\documentclass[aps,prl,onecolumn,superscriptaddress, 11pt]{revtex4-2}
\usepackage{graphicx}
\usepackage{amsmath}
\usepackage{amssymb}
\usepackage{textcomp}
\usepackage{url}
\usepackage{epsfig}

\usepackage{multirow}
\usepackage{siunitx}
\usepackage{color}

\usepackage[colorlinks,
linkcolor=blue,
citecolor=blue,
urlcolor=blue]{hyperref}

\begin{document}

\bibliographystyle{naturemag}

\title{Efficient generation and amplification of intense vortex and vector laser pulses via strongly coupled stimulated Brillouin scattering in plasmas}



\author{Yipeng Wu}
\email{wuyipeng@ucla.edu}
\affiliation{University of California, Los Angeles, California 90095, USA}
\author{Chaojie Zhang}
\affiliation{University of California, Los Angeles, California 90095, USA}
\author{Zan Nie}
\affiliation{University of California, Los Angeles, California 90095, USA}
\author{Mitchell Sinclair}
\affiliation{University of California, Los Angeles, California 90095, USA} 
\author{Audrey Farrell}
\affiliation{University of California, Los Angeles, California 90095, USA}
\author{Kenneth A Marsh}
\affiliation{University of California, Los Angeles, California 90095, USA}
\author{E. Paulo Alves}
\affiliation{University of California, Los Angeles, California 90095, USA}
\author{Frank Tsung}
\affiliation{University of California, Los Angeles, California 90095, USA}
\author{Warren B. Mori}
\affiliation{University of California, Los Angeles, California 90095, USA}
\author{Chan Joshi}
\email{cjoshi@ucla.edu}
\affiliation{University of California, Los Angeles, California 90095, USA}


\begin{abstract}
The past decade has seen tremendous progress in the production and utilization of vortex and vector laser pulses. 
Although both are considered as structured light beams, 
the vortex lasers have helical phase fronts and phase singularities, while the vector lasers have spatially variable polarization states and polarization singularities. In contrast to the vortex pulses that carry orbital angular momentum (OAM), the vector laser pulses have a complex spin angular momentum (SAM) and OAM coupling. Despite many potential applications enabled by such pulses, the generation of high-power/-intensity vortex and vector beams remains challenging. Here, we demonstrate using theory and three-dimensional simulations that the strongly-coupled stimulated Brillouin scattering (SC-SBS) process in plasmas can be used as a promising amplification technique with up to 65$\%$ energy transfer efficiency from the pump beam to the seed beam for both vortex and vector pulses. 
We also show that SC-SBS is strongly polarization-dependent in plasmas, enabling an all-optical polarization control of the amplified seed beam.
Additionally, the interaction of such structured lasers with plasmas leads to various angular momentum couplings and decouplings that produce intense new light structures with controllable OAM and SAM.
This scheme paves the way for novel optical devices such as plasma-based amplifiers and light field manipulators.
\end{abstract}


\maketitle

\section*{Introduction}

Light can possess orbital angular momentum (OAM) depending on its phase structure, and spin angular momentum (SAM) depending on its polarization state.  
In recent years, much attention has been focused on structured light \cite{NP_Structured_Light, Light_Optical_Vorticies, LPR_Yijie} 
where one can arbitrarily tailor light in all its degrees of freedom, especially phase and polarization.
The most familiar example of structured light is the vortex beam, exemplified by the Laguerre-Gaussian (LG) modes carrying well-defined OAM \cite{Allen_1992, Light_Optical_Vorticies}.
These modes have a helical phase structure with a phase singularity at the beam center.
Another type of structured light of particular interest is the vector beam, also referred to as the Poincaré beam \cite{Light_Optical_Vorticies, OE_Holleczek, Khajavi_2016, LPR_Yijie, PRL_Poincare, Rosales_2018},
which is characterized by space-variant polarization state and has polarization singularities.
The simplest example is a radially or azimuthally polarized cylindrical vector beam \cite{OE_Holleczek, AOP_Zhan} where the electric field either points in the radial or azimuthal directions for all angles.
In contrast to the vortex beams carrying OAM, the vector beams are characterized by complex SAM-OAM coupling.
These two types of beams can be described by the high-order Poincaré Sphere (HOPS) \cite{NP_Structured_Light, PRL_Poincare, PRL_Milione, Rosales_2018}.
The two poles of the HOPS denote orthogonal circularly polarized LG vortex beams with different topological charges, which carry both SAM and OAM. 
The other points on the HOPS describe the spatially variable polarization fields of vector beams, which can be represented by linear superpositions of the two poles \cite{PRL_Poincare, PRL_Milione, Rosales_2018}. 

The vortex and vector beams have opened new opportunities in a wide range of fields, 
such as 
high-resolution microscopy \cite{OL_Chen}, 
optical manipulation \cite{Nature_optical_manipulation, NP_manipulation} and optical communication \cite{NP_optical_communication, PR_Wang}.
At intensities beyond solid material breakdown thresholds, the applications of both laser beams have also been largely explored theoretically, including magnetic field generation in plasmas \cite{Shi_2018, PRE_Nuter, Sederberg_2020, Longman_PRR, Yipeng_PRR}, 
compact laser-driven particle acceleration \cite{OE_Wong, HPLSE, Vieira_PRL2, Vieira_PRL, PRL_Marceau, PRL_Zaim, Shi_PRL_DLA, OE_Wen}, and intense high-order harmonic generation \cite{PRL_Vortex_HHG, PRA_Chen}.
However, experimental demonstrations of these schemes have been hindered due to the difficulty of generating such high-intensity/-power beams.
In low-intensity/-power applications, the vortex and vector beams can be generated by using spiral phase plates \cite{Beijersbergen1994}, $q$-plates \cite{PRL_Marrucci}, computer generated holograms \cite{OQE_Heckenberg}, cylindrical-lens mode convertors \cite{OC_Beijersbergen},
and spin–orbit optics \cite{PRL_Brasselet, Science_Devlin}.
However, most such methods cannot be used in high-intensity/power laser systems due to their low damage thresholds and nonlinearities introduced by transmitting ultrashort laser pulses through optical materials.

Since a fully-ionized plasma can sustain extremely high electromagnetic fields,
the use of a plasma as the optical medium is a promising solution.
One feasible way is to first generate a weak vortex/vector pulse utilizing the above-mentioned conventional optics and then amplify it to high intensity/power in a plasma.
Such amplification relies on a three-wave coupling where a plasma wave is used as an optical component to scatter the energy from a long pump pulse to a short seed pulse.
The plasma wave can be either a high-frequency Langmuir electron wave as in stimulated Raman scattering (SRS, also called Raman amplification) \cite{PRL_Raman_first, PRL_Malkin, Raman_exp, NP_Trines, PRL_Trines, PRL_Mendon} or a low-frequency ion-acoustic wave (IAW) as in stimulated Brillouin scattering (SBS, also called Brillouin amplification) \cite{Milroy_1977, POF_1979, POP_Andreev}.
Recently, Vieira et al. have shown through simulations that SRS can be used to amplify vortex lasers in plasmas \cite{NC_Vieira}.
Nevertheless, this scheme to-date has not been suggested for 
manipulation of polarization states
and generation/amplification of vector beams.
In addition, SRS is sensitive to parameter fluctuations.
It requires a relatively uniform and long (mm to cm scale) plasma to satisfy the frequency matching condition and provide sufficient gain (amplification),
which is difficult to achieve in practice.
Moreover, SRS is also affected by kinetics effects \cite{PRL_Malkin}.

In this paper we propose that greater experimental robustness and higher pump to seed amplification efficiency can be achieved by utilizing strongly coupled Brillouin amplification process (SC-SBS).
Historically Brillouin amplification was first considered for plasma-based pulse amplification in the weak coupling regime (WC-SBS) over forty years ago \cite{Milroy_1977, POF_1979}.
However, the duration of the amplified pulse in this regime is relatively long ($>$ps), which is set by the period of IAW.
Later studies have shown that significant improvement can be obtained when Brillouin amplification is transitioned from the WC-SBS regime to the SC-SBS regime \cite{POP_Andreev},
where at larger pump intensities, 
the ion-plasma response time can be decreased. 
It is therefore suitable for amplifying shorter (sub-ps) seed pulses.
Additionally, in SC-SBS, the IAW frequency is much smaller than the pump frequency.
Hence, the frequency matching condition does not depend substantially on the local plasma density and can be automatically satisfied even though the pump and the seed are wavelength-degenerate. 
Since little photon energy is lost to the IAW, the seed can efficiently carry off most of the incident pump energy.
Furthermore, the energy transfer from the pump to the seed is very fast, allowing for short amplifier lengths (sub-mm).
In the past two decades, much progress has been achieved in understanding and optimizing SC-SBS \cite{POP_Andreev, PRL_Lancia2, PRL_Lancia, PRL_Weber, PRL_SBS_chirp, POP_Chiaramello, POP_Lehmann, POP_Edwards,  POP_Amiranoff, PRX_experiment, SR_Trines, Alves_2021}. 
Very recently, laser amplification above the Joule level through SC-SBS has been experimentally demonstrated with a large energy transfer efficiency up to $20\%$ \cite{PRX_experiment}.
However, all these studies have used Gaussian pump and seed beams.
The exploration of SC-SBS for vortex and vector lasers with tailored phase and polarization is still lacking.

Here, we explore how these additional degrees of freedom affect the SC-SBS process via theory and three-dimensional (3D) particle-in-cell (PIC) simulations. 
We show the matching rules of both OAM and polarization states during the SC-SBS process.
Based on these matching rules, we demonstrate the feasibility of simultaneously manipulating the OAM and polarization (e.g., linear, circular, elliptical polarizations) of a vortex beam while boosting its intensity/power.
We also show the coupling of OAM and SAM inside the plasma, leading to the generation and amplification of a new vector laser pulse from a vortex seed beam.
Furthermore, we also demonstrate that SC-SBS process can decouple SAM and OAM, and then create and intensify a new vortex beam from a vector seed pulse.

\section*{Results}
\subsection{Theoretical analysis}

Assuming that the pump and the seed propagate in a homogeneous plasma along the $+z$ and $-z$ directions, respectively, 
the general coupled three-wave envelope equations describing the amplitude evolution of the pump, the seed, and the IAW in the SC-SBS process are given by 
(see Supplementary Note 1 for detailed derivations)
\begin{equation}\label{eq1}
\begin{aligned}
(\frac{\partial}{\partial t}+v_{\rm{g}}^{\rm{P}}\frac{\partial}{\partial z}-i\frac{c^2}{2\omega_{\rm{P}}}\nabla_\perp^2)\vec{E_{\rm{P}}}=-i\frac{\omega_{\rm{pe}}^2}{4\omega_{\rm{S}}}\delta N\vec{E_{\rm{S}}}
\end{aligned}
\end{equation}
 \begin{equation}\label{eq2}
\begin{aligned}
(\frac{\partial}{\partial t}-v_{\rm{g}}^{\rm{S}}\frac{\partial}{\partial z}-i\frac{c^2}{2\omega_{\rm{S}}}\nabla_\perp^2)\vec{E_{\rm{S}}}=-i\frac{\omega_{\rm{pe}}^2}{4\omega_{\rm{P}}}\delta N^*\vec{E_{\rm{P}}}
\end{aligned}
\end{equation}
\begin{equation}\label{eq3}
\begin{aligned}
(\frac{\partial^2}{\partial t^2}-v_{\rm{IAW}}^2\frac{\partial^2}{\partial z^2}-v_{\rm{IAW}}^2\nabla_\perp^2)\delta N=-\frac{Ze^2k_{\rm{IAW}}^2}{2m_{\rm{e}}m_{\rm{i}}\omega_{\rm{P}}\omega_{\rm{S}}}(\vec{E_{\rm{P}}}\cdot \vec{E_{\rm{S}}}^*)
\end{aligned}
\end{equation}

Equations \eqref{eq1} and \eqref{eq2} describe the laser propagation in the plasma,
where the subscript or superscript $\rm{P}$ and $\rm{S}$ denote quantities of the pump and the seed, respectively.
$\vec{E_j}$ ($j=\rm{P,S}$) is the electric field envelope of the pump/seed laser, with the electric field given by $\frac{1}{2}\vec{E_j}(x,y,z,t)\exp[i(\pm k_jz-\omega_jt)]+\frac{1}{2}\vec{E_j}^*(x,y,z,t)\exp[-i(\pm k_jz-\omega_jt)]$, where $\omega_j$ is the laser frequency, $k_j$ is the laser wavenumber in the plasma, the superscript $*$ refers to the complex conjugate, and the plus and minus signs ($\pm$) before $k_j$ indicate that the two laser pulses travel in opposite directions.
$v_{\rm{g}}^j=k_jc^2/\omega_j$ is the laser group velocity in the plasma.
Each laser also obeys the linear dispersion relation of electromagnetic waves in the plasma, 
i.e., $c^2k_{j}^2=\omega_{j}^2-\omega_{\rm{pe}}^2$,
where $c$ is light speed in vacuum, $\omega_{\rm{pe}}=\sqrt{\frac{n_{\rm{e}0} e^2}{\epsilon_0 m_{\rm{e}}}}$ is the electron plasma frequency, $n_{\rm{e}0}$ is the background electron density, $\epsilon_0$ is the vacuum electric permittivity, $e$ is the elementary charge and $m_{\rm{e}}$ is the electron mass.

Equation \eqref{eq3} describes the grating-like electron density perturbation associated with the IAW excited by the beating of the pump pulse with the seed pulse.
The relative electron density perturbation ($\frac{\delta n_{\rm{e}}}{n_{\rm{e}0}}$)
is expressed as $\frac{1}{2}\delta N(x,y,z,t)\exp[i(k_{\rm{IAW}}z-\omega_{\rm{IAW}}t)]+\frac{1}{2}\delta N^*(x,y,z,t)\exp[-i(k_{\rm{IAW}}z-\omega_{\rm{IAW}}t)]$, where $\delta N$ is the relative density perturbation envelope, $\omega_{\rm{IAW}}$ is the IAW frequency, $k_{\rm{IAW}}\approx k_{\rm{P}}+k_{\rm{S}}$ is the IAW wavenumber, $v_{\rm{IAW}}=\sqrt{ZT_{\rm{e}}/m_{\rm{i}}}$ is the ion acoustic velocity for negligible ion temperature,
$Z$ is the ion charge, $m_{\rm{i}}$ is the ion mass, and $T_{\rm{e}}$ is the electron temperature.
The dot product $(\vec{E_{\rm{P}}}\cdot \vec{E_{\rm{S}}}^*)$ indicates that the generation of the electron density perturbation or IAW inherently depends on the polarization states of the pump and the seed, 
being maximum for identical polarizations and zero for orthogonal polarizations. 
As shown in Eqs. \eqref{eq1} and \eqref{eq2}, the evolutions of the pump and the seed are coupled with the density perturbation or IAW, 
therefore the energy transfer from the pump to the seed is also polarization-dependent. 

In SC-SBS,
the pump and seed lasers can initially have the same wavelength so that $\omega_{\rm{P}}= \omega_{\rm{S}}= \omega_0$, $k_{\rm{P}}= k_{\rm{S}} = k_0=\frac{1}{c}\sqrt{\omega_0^2-\omega_{\rm{pe}}^2}$, $v_{\rm{g}}^{\rm{P}}= v_{\rm{g}}^{\rm{S}}= v_{\rm{g}}=k_0c^2/\omega_0$ and $k_{\rm{IAW}} \approx 2k_0$.
For simplicity, we assume that the pump and seed are linearly polarized in the same direction, 
and consider that the electric field envelope of each laser $\vec{E_j}$ ($j=\rm{P,S}$) can be written as $\vec{E_j}(x,y,z,t)=\vec{E_{||,j}}(z,t)T_j(x,y,z)$, where $\vec{E_{||,j}}$ represents the longitudinal envelope profile (a function of $t$ and $z$) and $T_j$ represents the transverse envelope profile (a function of $x$, $y$, and $z$).
Similarly, the density perturbation envelope $\delta N$ can be expressed as $\delta N(x,y,z,t)=\delta N_{||}(z,t)T_{\delta N} (x,y,z)$, with $\delta N_{||}$ the longitudinal profile and $T_{\delta N}$ the transverse profile.
By assuming that $T_{j}$ obeys the paraxial wave equation,
we can simplify Eqs. \eqref{eq1}-\eqref{eq3} to obtain
\begin{equation}\label{eq4}
\begin{aligned}
T_{\rm{P}}(\frac{\partial}{\partial t}+v_{\rm{g}}\frac{\partial}{\partial z})\vec{E_{||,\rm{P}}}=-i\frac{\omega_{\rm{pe}}^2}{4\omega_0}  T_{\delta N}T_{\rm{S}}\delta N_{||}\vec{E_{||,\rm{S}}}
\end{aligned}
\end{equation}
 \begin{equation}\label{eq5}
\begin{aligned}
T_{\rm{S}}(\frac{\partial}{\partial t}-v_{\rm{g}}\frac{\partial}{\partial z})\vec{E_{||,\rm{S}}}=-i\frac{\omega_{\rm{pe}}^2}{4\omega_0} T_{\delta N}^* T_{\rm{P}}\delta N_{||}^*\vec{E_{||,\rm{P}}}
\end{aligned}
\end{equation}
\begin{equation}\label{eq6}
\begin{aligned}
T_{\delta N}(\frac{\partial^2}{\partial t^2}-v_{\rm{IAW}}^2\frac{\partial^2}{\partial z^2})\delta N_{||}=-\frac{2Ze^2k_0^2}{m_{\rm{e}}m_{\rm{i}}\omega_0^2}T_{\rm{P}}T_{\rm{S}}^*\vec{E_{||,\rm{P}}}\cdot \vec{E_{||,\rm{S}}}^* 
\end{aligned}
\end{equation}


Next we consider a LG vortex pump or seed beam.
The LG modes are normally indexed by two mode numbers, denoted as their radial and azimuthal indexes. 
For simplicity, here we consider that the radial mode index is zero (the case of nonzero radial mode will be explored in our future publications) and the interaction length is much smaller than its Rayleigh length,
then the transverse envelope profile $T_j$ ($j=\rm{P,S}$) can be written as $T_j(x,y,z)\approx T_{r,j}(x,y)\exp(il_j\phi)$, 
where $T_{r,j}(x,y)\approx \left(\frac{\sqrt{2}r}{W_{0,j}}\right)^{|l_j|}\exp\left(-\frac{r^2}{W_{0,j}^2}\right)$. 
Here, 
$r=\sqrt{x^2+y^2}$ is the radial distance to the axis, $W_{0,j}$ is the laser spot size at the focal plane, $l_j$ is the azimuthal mode index (also known as the topological charge), 
and $\phi=\arctan(y/x)$ is the azimuthal angle.
Clearly, the helical azimuthal phase term of $\exp(il_j\phi)$ is presented.
Similarly, the transverse envelope of the electron density perturbation $T_{\delta N}$ is given by $T_{\delta N}(x,y,z)\approx T_{r,\delta N} (x,y)\exp(il_{\delta N}\phi)$,
where $l_{\delta N}$ corresponds to the plasmon OAM state of IAW.

Equations \eqref{eq4}-\eqref{eq6} suggest a matching of transverse 
envelope profile of these waves.
The azimuthal phase term of the left side in Eqs. \eqref{eq4}-\eqref{eq6} must be equal to that of the right side, i.e., $\exp(il_{\rm{P}}\phi)=\exp(il_{\rm{S}}\phi)\exp(il_{\delta N}\phi)$ must be satisfied,
leading to $l$-number matching condition or OAM conservation equation for the pump, seed and IAW, given by $l_{\rm{P}}=l_{\rm{S}}+l_{\delta N}$.

By cancelling the azimuthal phase terms in Eqs. \eqref{eq4}-\eqref{eq6} 
and working out the dispersion relation, 
we can obtain the IAW frequency of $\omega_{\rm{IAW}}=\omega_{\rm{sc}}=\frac{1}{2}\left(\beta_{\rm{g}}^2\omega_0 \omega_{\rm{pi}}^2{a_{\rm{P0}}}^2\right)^{\frac{1}{3}}$ and the seed growth rate of $\gamma_{\rm{sc}}=\frac{\sqrt{3}}{2}\left(\beta_{\rm{g}}^2\omega_0 \omega_{\rm{pi}}^2{a_{\rm{P0}}}^2\right)^{\frac{1}{3}}$, where $\beta_{\rm{g}}=v_{\rm{g}}/c=\sqrt{1-\omega_{\rm{pe}}^2/\omega_0^2}=\sqrt{1-n_{\rm{e0}}/n_{\rm{c}}}$ is the normalized pump/seed group velocity with $n_{\rm{c}}$ the critical density above which the pump/seed can no longer propagate, $a_{\rm{P}}=\frac{e|\vec{E_{||,\rm{P}}}|T_{r,\rm{P}}}{m_{\rm{e}}\omega_0c}$ is the normalized pump vector potential with $a_{\rm{P0}}=a_{\rm{P}}(t=0)$ referring to the initial incident value, and $\omega_{\rm{pi}}=\sqrt{\frac{n_{\rm{i0}}Z^2e^2}{\epsilon_0 m_{\rm{i}}}}$ is the ion plasma frequency with $n_{\rm{i0}}=n_{\rm{e0}}/Z$ the unperturbed ion density.
Here we note that 
$\gamma_{\rm{sc}}$ varies in different transverse positions since $\gamma_{\rm{sc}}\propto a_{\rm{P0}}^{2/3}$ and $a_{\rm{P0}}$ is $r$-dependent.
We also note that the strong coupling limit is characterized by
$\left(\frac{v_{\rm{osc}}}{v_{\rm{e}}}\right)^2>4k_{\rm{P}}v_{\rm{IAW}}\frac{\omega_{\rm{P}}}{\omega_{\rm{pe}}^2}$ \cite{POP_Andreev},
where 
$v_{\rm{osc}}=a_{\rm{P}0}c$ 
and $v_{\rm{e}}=\sqrt{T_{\rm{e}}/m_{\rm{e}}}$. 
In this limit, the IAW associated with the density perturbation $\delta N$ grows significantly over one IAW period. In practical units, the above equation can be rewritten as
\begin{equation}\label{eq8}
\begin{aligned}
I_{\rm{P}0}^{\rm{peak}}[{\rm{W/cm^{2}}}]\lambda_{\rm{P}}^2[\mu \rm{m}]>1.1\times10^{13}T_e^{3/2}[{\rm keV}]\times \frac{n_c}{n_{e0}}\sqrt{1-n_c/n_{e0}}
\end{aligned}
\end{equation}
where $I_{\rm{P}0}[{\rm{W/cm^{2}}}]=1.3\times 10^{18}{a_{\rm{P}0}}^2/\lambda_{\rm{P}}^2[\mu \rm{m}]$ is the incident pump laser intensity, $\lambda_P$ is the pump wavelength, and the superscript ``peak" denotes the peak value.

The SC-SBS process
happens in three main stages \cite{PRL_SBS_chirp,  POP_Amiranoff}.
The first stage is the ``initial" stage where the pump starts to interact with the seed and creates the density perturbation, allowing the phases of the pump, seed and density perturbation to adapt in order to start the energy transfer.
The second stage is know as the ``linear'' stage corresponding to the so-called linear SC-SBS solution, during which the pump depletion is negligible and the seed grows exponentially as 
$a_{\rm{S}}(t)\propto \exp({\gamma_{\rm{sc}}t})$ \cite{POP_Chiaramello, POP_Amiranoff},
where $a_{\rm{S}}=\frac{e|\vec{E_{||,\rm{S}}}|T_{r,\rm{S}}}{m_{\rm{e}}\omega_0c}$ is the seed normalized vector potential.
This exponential growth is also accompanied by a temporal stretching of the seed pulse if its initial duration is short enough. 
The second stage ends when $a_{\rm{S}}(t) \simeq a_{\rm{P0}}$ and the SC-SBS will enter the last ``self similar" stage \cite{POP_Andreev, PRL_Lancia} where the pump depletion is non-negligible 
and the seed scales as
$a_{\rm{S}}(t)\propto a_{\rm{P0}} (\gamma_{\rm{sc}}t)^\alpha$,
with $0.75 \leq  \alpha \leq0.9$ \cite{POP_Andreev, SR_Trines, Alves_2021}.
In addition to this power-law amplitude growth, 
the seed will also be temporally compressed down to ${\gamma_{\rm{sc}}^{\rm{peak}}}^{-1}$ \cite{POP_Andreev}.

The above analysis is based on the assumption that the pump and seed pulses are linearly polarized in the same direction. Now we extend the results to a more general polarization case,
where the laser polarization state is described in terms of two orthogonal polarization bases, e.g., the horizontal ($x$) and vertical ($y$) linear polarization bases.
The generic expression for the electric field envelope of the pump ($j=\rm{P}$) or seed ($j=\rm{S}$) is 
\begin{equation}\label{eq9}
\begin{aligned}
&\vec{E_j}=E_{j,x}\vec{e_x}+\exp(i\psi_j^{x,y})E_{j,y}\vec{e_y}
\end{aligned}
\end{equation}
where 
$E_{j,x}\propto a_{j,x}\exp(il_{j,x}\phi)$ and $E_{j,y}\propto a_{j,y}\exp(il_{j,y}\phi)$ represent the electric field envelope along the direction given by the horizontal and vertical unitary vectors $\vec{e_x}$ and $\vec{e_y}$, respectively.
$\psi_j^{x,y}$ is the relative phase shift between the $E_{j,x}$ and $E_{j,y}$ components.
In each polarization plane, the electric field components of both the pump and seed should satisfy the above three-wave equations and also the OAM matching condition, 
leading to 
\begin{equation}\label{eq10}
\begin{aligned}
l_{{\rm{P}},x}=l_{{\rm{S}},x}+l_{\delta N},\ 
l_{{\rm{P}},y}=l_{{\rm{S}},y}+l_{\delta N}
\end{aligned}
\end{equation}

We note that apart from the OAM matching, the polarization states of the pump and the seed can also be ``matched" since the SC-SBS gain is strongly polarization-dependent.
As we shall see in the following simulations, 
if the pump and the seed beams have different initial polarization states,
the output polarization state of the amplified seed pulse could be 
quasi-identical
to the input pump polarization.
The degree of polarization matching depends on the SC-SBS gain in different polarization planes.
This promising and unique feature can allow polarization control of the vortex seed beam by tuning the polarization state of the pump beam,
which complements the suite of existing plasma-mediated polarization manipulation schemes for high-intensity/-power Gaussian pulses \cite{polarization_PRL1, polarization_PRL2, polarization_PRL3, polarization_PRL4, polarization_PRE1, polarization_PRE2, polarization_MRE}.

In addition to linear polarization bases, orthogonal circular polarization bases with different handedness (i.e., right- and left-handed) can also be used to describe an arbitrary polarization state.
In this case, the electric field envelope of the pump ($j=\rm{P}$) or seed ($j=\rm{S}$) is
\begin{equation}\label{eq11}
\begin{aligned}
&\vec{E_j}=E_{j,\rm{R}}\vec{e_{\rm{R}}}+\exp(i\psi_j^{\rm{R,L}})E_{j,\rm{L}}\vec{e_{\rm{L}}}
\end{aligned}
\end{equation}
where 
$E_{j,\rm{R}}\propto a_{j,\rm{R}}\exp(il_{j,\rm{R}}\phi)$ and $E_{j,\rm{L}}\propto a_{j,\rm{L}}\exp(il_{j,\rm{L}}\phi)$ represent the electric field envelope along the direction given by the right- and left-handed circular unitary vectors $\vec{e_{\rm{R}}}=(\vec{e_x}-i\vec{e_y})/\sqrt{2}$ and $\vec{e_{\rm{L}}}=(\vec{e_x}+i\vec{e_y})/\sqrt{2}$, respectively.
$\psi_j^{\rm{R,L}}$ is the relative phase shift between $E_{j,\rm{R}}$ and $E_{j,\rm{L}}$.
Note that vector beams with space-variant polarization states are a linear superposition of orthogonal circularly polarized vortex beams with different OAM modes.
Therefore, the electric field envelope of a vector pump or seed can be exactly written as the above Eq. \eqref{eq11} with $l_{j,\rm{R}}\neq l_{j,\rm{L}}$ (see Supplementary Note 2 for details).
Similar to the case of linear polarization bases,
we can obtain the OAM conservation equation in two circular polarization bases, given by 
\begin{equation}\label{eq12}
\begin{aligned}
l_{\rm{P,R}}=l_{\rm{S,R}}+l_{\delta N},\ 
l_{\rm{P,L}}=l_{\rm{S,L}}+l_{\delta N}
\end{aligned}
\end{equation}
Additionally, the polarization matching condition between the pump and the seed can also be achieved in this case.
These matching rules of both OAM and polarization are the basis for the coupling and/or decoupling of SAM and OAM, including the creation of new vector modes, as we have shown later in this paper.

\subsection{3D PIC simulations.}
To illustrate our scheme, we perform a series of 3D PIC simulations with different pump and seed configurations using the code OSIRIS \cite{Fonseca_2002}.
In these simulations, the variable parameters are laser modes, spot sizes and peak intensities of both the pump and seed beams.
All other parameters are identical.
The pump has a flat-top temporal profile with FWHM duration of 800 fs and the seed has a sin$^2$ temporal profile with FWHM duration of 100 fs.
Both pulses have the same wavelength of 
$\lambda_{\rm{P}}=\lambda_{\rm{S}}=2\pi c/\omega_0=1 \mu$m.
As mentioned earlier,
the degenerate frequencies of the pump and the seed still leads to SC-SBS amplification because the bandwidth of the short seed pulse ($>10$ THz) is greater than the IAW frequency.
The background plasma (hydrogen) is composed of electrons and ions with charge of $Z=1$, mass of $m_{\rm{i}}=1836m_{\rm{e}}$, electron temperature of $T_{\rm{e}}= 0.5$ keV, and ion temperature of $T_{\rm{i}}=0.01$ keV. 
The plasma has a plateau density profile with length of $500c/\omega_0=80\ \mu$m and electron density of $n_{\rm{e0}}=0.3n_{\rm{c}}=3.34\times10^{20}$cm$^{-3}$.
Since the plasma density is greater than $0.25n_{\rm{c}}$, the pump cannot go unstable via the SRS instability.
The pump meets the seed at the right boundary of the plasma.
All the simulations utilize pump intensities significantly above the strong coupling threshold,
which corresponds to $1.1\times10^{13}$ W/cm$^2$ for the above plasma conditions [according to Eq. \eqref{eq8}].
We use the fixed window configuration in modeling the self-consistent dynamics of both the pump and seed pulses, 
which captures the full propagation of these two beams in the plasma, including the thermal filamentation instability \cite{POF_Kaw, PRL_Max, PRL_Perkins, PRL_Epperlein} that may arise before the pump 
encounters the seed.

\begin{figure}[t]
\centering\includegraphics[height=0.46\textwidth]{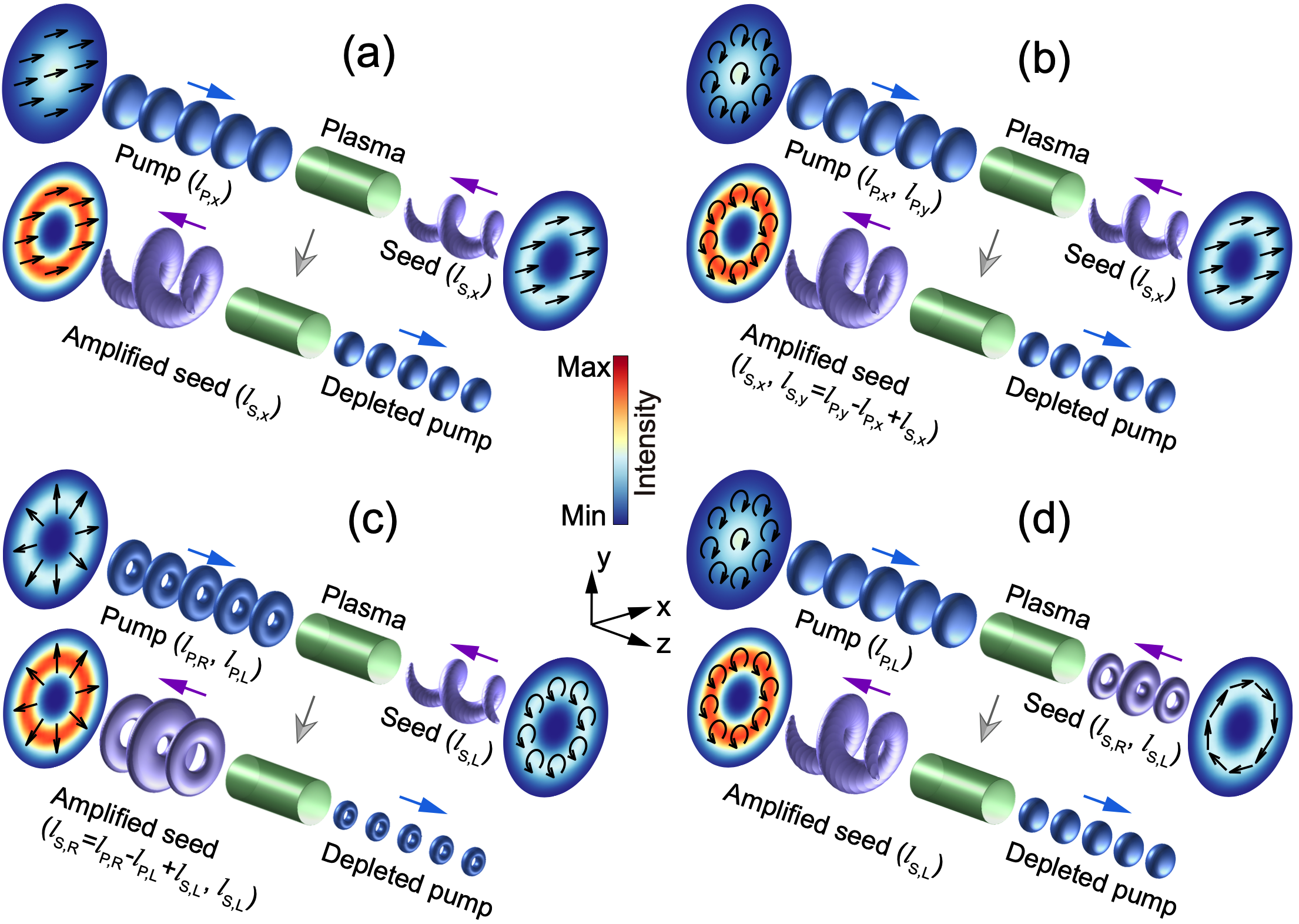}
\caption{\label{fig1} 
Schematic diagram for the generation and amplification of vortex and vector beams via SC-SBS. 
The pump/seed lasers propagate in the direction of the blue/purple arrow, respectively.
The plasma is shown by the green cylinder.
The back/front projections show the intensity profile and polarization pattern of the closest laser.
(a) Amplification of vortex beams by Gaussian or LG pump beams. 
(b) Generation and amplification of vortex beams with new OAM modes and controllable polarization. 
(c) Generation and amplification of vector beams from vortex seed beams. 
(d) Generation and amplification of vortex beams from vector seed beams. 
}
\end{figure}

Note that the growth rate for SC-SBS increases with an increase in the plasma density ($\gamma_{\rm{sc}}\propto (\beta_{\rm{g}}^2\omega_{\rm{pi}}^2)^{1/3}\propto [(1-n_{\rm{e0}}/n_{\rm{c}})n_{\rm{e0}}/n_{\rm{c}}]^{1/3}$).
Here we choose a high plasma density ($n_{\rm{e0}}=0.3n_{\rm{c}}$) to obtain fast growth of the Brillouin-amplified pulse in a short (80 $\mu$m) plasma.
Therefore the simulation box length and simulation time can be reduced to acceptable values.
We also use moderate pump/seed spot sizes ($\sim$10 $\mu$m scale) to reduce the simulation box width, making full 3D simulations numerically feasible. 
It is straightforward to extend these simulation results to low plasma density cases and large spot sizes.
Based on the above arguments, the simulation boxes are chosen to have a dimension of 660$\times$350$\times$350$(c/\omega_0)^3$ (105$\times$55.7$\times$55.7$\mu$m$^3$) in the $z$, $x$ and $y$ directions, respectively. 
The plasma is located between $z=150 c/\omega_0$ and $z=650 c/\omega_0$.
The longitudinal regions where $0\leq z \leq 150c/\omega_0$ and $650c/\omega_0 \leq z\leq 660 c/\omega_0$ are vacuum.
The pump enters the simulation box from the left boundary ($z=0$) at $t=0$ and the seed enters from the right ($z=660 c/\omega_0$) at $t=700 \omega_0^{-1}$.
We will now discuss in detail the four cases shown in Fig. \ref{fig1}.

\subsubsection{}
\textbf{Case 1: Amplification of vortex beams by Gaussian or LG pump beams.}
Figure \ref{fig1}(a) illustrates this case, where a linearly polarized (assuming in the $x$ direction) LG seed carrying OAM mode of $l_{\rm{S},x}$
is amplified in a plasma by a counter-propagating pump with the same polarization state.
The pump can be either a Gaussian beam without OAM ($l_{{\rm{P}},x}=0$) or a LG vortex beam carrying OAM ($l_{{\rm{P}},x}\neq 0$).
According to $l$-number matching condition, when such a pump $\vec{E_{\rm{P}}}=E_{{\rm{P}},x}\vec{e_x}\propto a_{\rm{P}}\exp(il_{{\rm{P}},x}\phi )\vec{e_x}$ interacts with the vortex seed $\vec{E_{\rm{S}}}=E_{{\rm{S}},x}\vec{e_x}\propto a_{\rm{S}}\exp(il_{{\rm{S}},x}\phi )\vec{e_x}$ in the plasma, 
an electron density perturbation of
$\delta N \sim \exp (il_{\delta N}\phi)$ with $l_{\delta N}=l_{{\rm{\rm{P}}},x}-l_{{\rm{S}},x}$ 
is generated.
Therefore, the IAW associated with this electron density perturbation also carries OAM mode of $l_{{\rm{P}},x}-l_{{\rm{S}},x}$.
The pump then beats with the IAW to further amplify the seed pulse.

\begin{figure}[t]
\centering\includegraphics[height=0.38\textwidth]{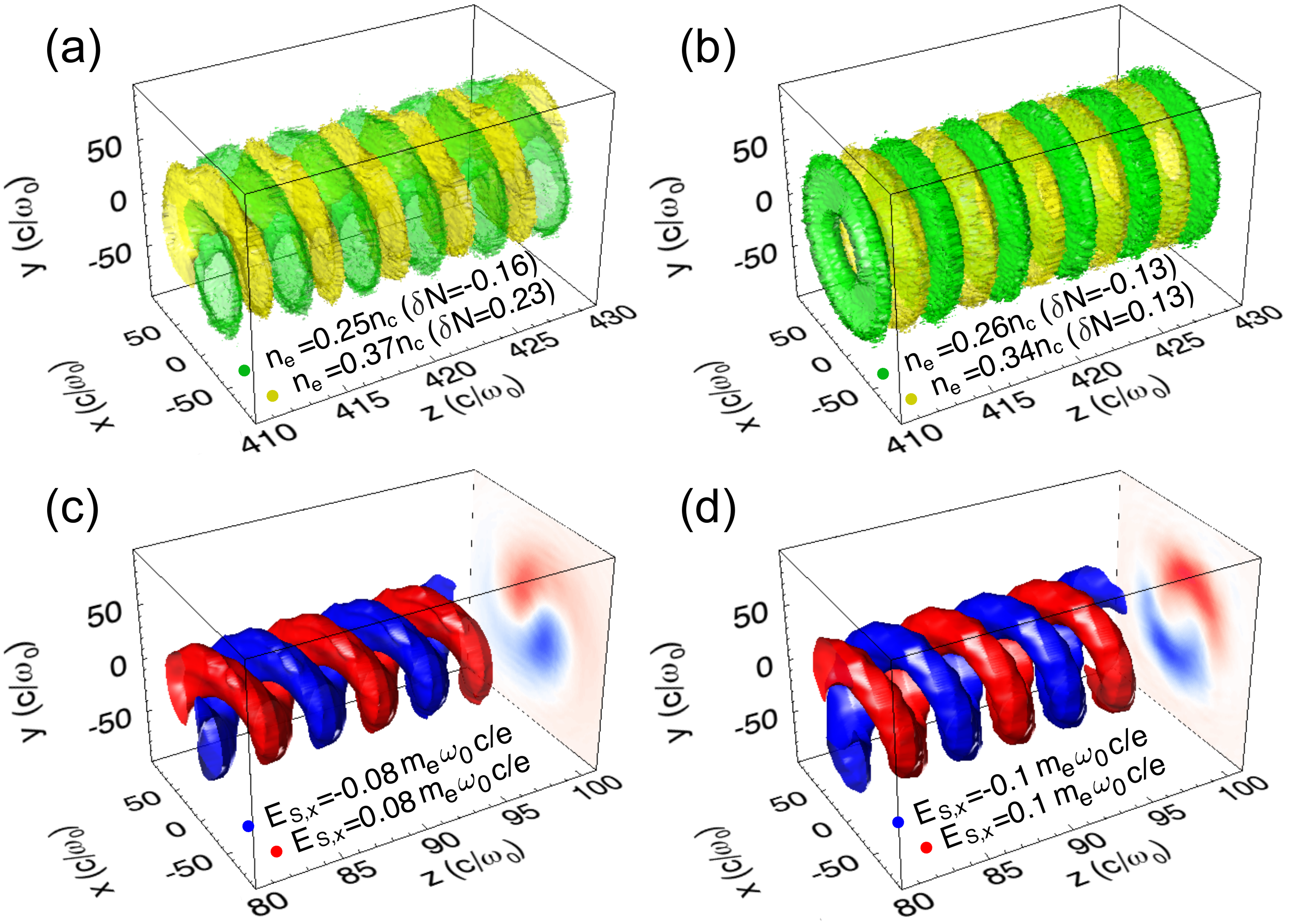}
\caption{\label{fig2} 
Simulations results showing the amplification of a $x-$linearly polarized LG vortex seed with $l_{{\rm{S}},x}=1$ by a $x-$linearly polarized Gaussian pump (left column) or LG vortex pump with $l_{{\rm{P}},x}=1$ (right column). The top row shows the 3D isosurfaces of the electron density perturbation in green and yellow colors. The bottom row  shows the 3D isosurfaces of the seed electric field (normalized to $m_e\omega_0c/e$) after amplification in blue and red colors.
Projections in the $x$-$y$ plane (blue–white–red) show the corresponding slice of the seed electric field at $z=80c/\omega_0$.
}
\end{figure}

Figure \ref{fig2} illustrates the simulation results showing the SC-SBS amplification of a $x-$linearly polarized LG seed pulse with $l_{{\rm{S}},x}=1$, focal spot size of $W_{0,\rm{S}}=12\ \mu$m, 
and initial peak intensity of $I_{\rm{S}0}^{\rm{peak}}=5\times10^{15}$ W/cm$^2$ (corresponding to peak vector potential of $a_{\rm{S}0}^{\rm{peak}}= 8.5\times 10^{-10}\sqrt{I_{\rm{S}0}^{\rm{peak}}[{\rm W/cm^{2}}]}\lambda_{\rm{S}}[\mu \rm{m}]=0.06$). 
The $x$-linearly polarized pump laser is either a Gaussian beam with $l_{{\rm{P}},x}=0$, focal spot size of $W_{0,{\rm{P}}}=15\ \mu$m (a bit larger than $W_{0,\rm{S}}$ in order for the seed to be contained by the pump volume) and incident peak intensity of $I_{\rm{P}0}^{\rm{peak}}=1\times10^{16}$ W/cm$^2$ ($a_{{\rm{P}}0}^{\rm{peak}}=0.085$),
or a LG vortex beam with $l_{{\rm{P}},x}=1$, $W_{0,\rm{P}}=12\ \mu$m and $I_{\rm{P}0}^{\rm{peak}}=1\times10^{16}$ W/cm$^2$ ($a_{\rm{P}0}^{\rm{peak}}=0.085$).
These values corresponds to $\gamma_{\rm{sc}}^{\rm{peak}}\approx 8\times10^{-3} \omega_0$.
For the Gaussian pump, the generated electron/ion density perturbation features a periodic (wavenumber of $k_{\rm{IAW}}\approx 2k_0$) helical structure with $l_{\delta N}=l_{{\rm{P}},x}-l_{{\rm{S}},x}=-1$ [Fig. \ref{fig2}(a)]; while for the LG vortex pump, the generated electron/ion density perturbation
has no azimuthal dependence
with $l_{\delta N}=l_{{\rm{P}},x}-l_{{\rm{S}},x}=0$ [Fig. \ref{fig2}(b)].
In both pump cases, the output amplified seed pulse retains its initial OAM with a helical structure,
as shown in Fig. \ref{fig2}(c) and \ref{fig2}(d).

\begin{figure}[t]
\centering\includegraphics[height=0.26\textwidth]{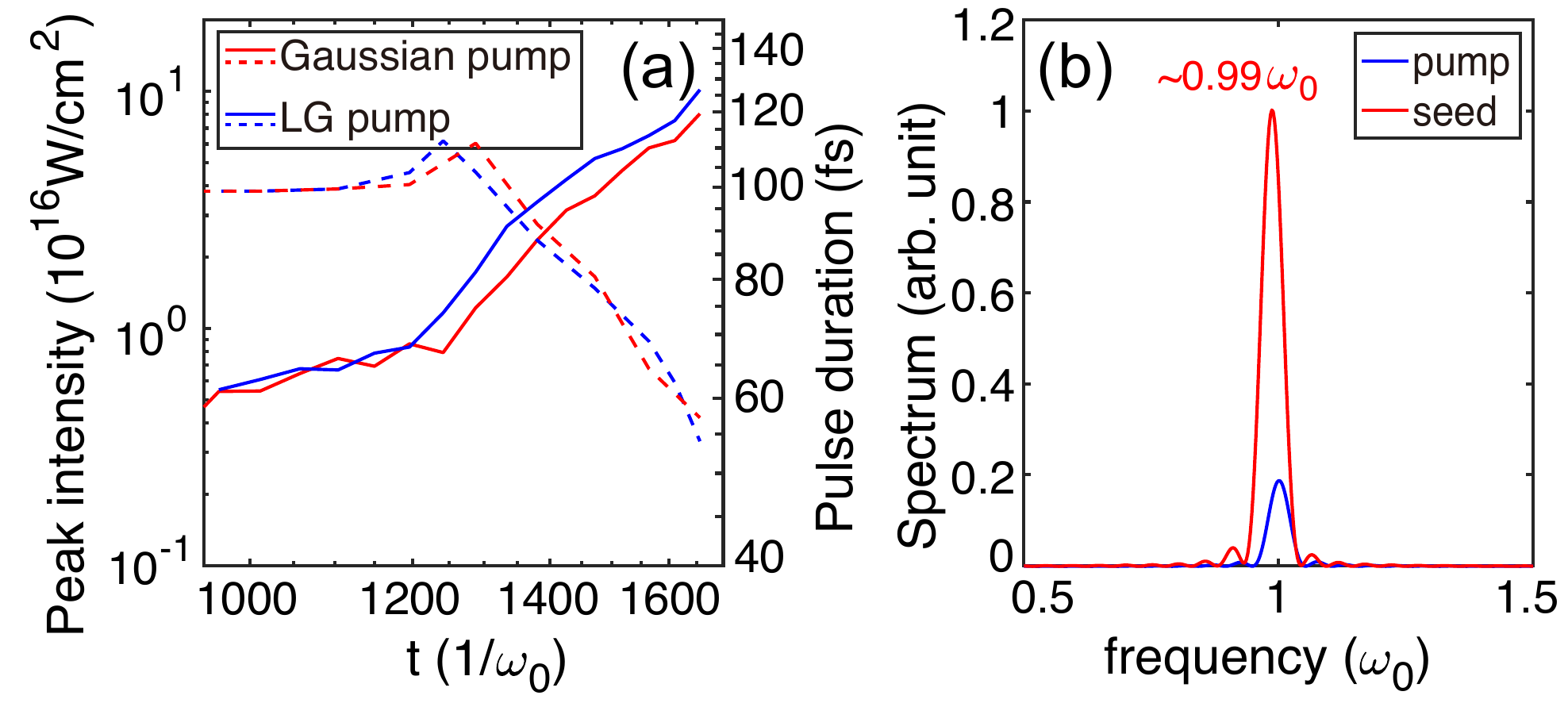}
\caption{\label{fig3} 
(a) Temporal evolution of peak seed intensity (solid lines) and seed FWHM duration (dashed lines) for both the Gaussian pump (red) and LG pump (blue) cases.
The electromagnetic field in the simulation box contains both the right-propagating pump and the left-propagating seed. 
We use $E_{{\rm{S}},x}=(E_x-\frac{1}{\beta} c B_y)/2$ to extract only the seed electric field from the simulation and then calculate the intensity, 
where $\beta$ is the laser group velocity normalized to $c$, with $\beta=1$ in vacuum and $\beta=\beta_{\rm{g}}=\sqrt{1-n_{\rm{e0}}/n_{\rm{c}}}$ in the plasma.
Note that all the axes are presented in the logarithmic scale. 
The seed intensity roughly increases as a power of time (linear in the logarithmic scale) in the ``self-similar" stage, in agreement with the theoretical prediction.
(b) Spectra of the amplified seed pulse (red) and the initial pump pulse (blue) in the LG pump case.
}
\end{figure}

Figure \ref{fig3}(a) shows the temporal evolution of the peak seed intensity and seed pulse duration
during the SC-SBS process.
After the seed overlaps with the pump, the amplification process first happens in the ``linear" stage ($t\lesssim1200\omega_0^{-1}$) where the seed grows in intensity and also stretches in duration from initial 100 fs to $\sim$115 fs.
Then for $t \gtrsim 1200\omega_0^{-1}$ the seed intensity is close to the pump intensity, the SC-SBS process mainly operates in the ``self-similar" pump-depletion stage, where there is efficient energy transfer from the pump to the seed. 
As mentioned earlier, this amplification is also accompanied by a temporal compression of the seed pulse.
The FWHM pulse duration has been compressed to about $110\omega_0^{-1}\approx58.3$ fs (Gaussian pump case) or $104\omega_0^{-1}\approx 55.1fs$ (LG pump case), which agrees well with the theoretical prediction of ${\gamma_{\rm{sc}}^{\rm{peak}}}^{-1}\approx 125 \omega_0^{-1}\approx 66.2$ fs.
Finally the seed pulse has been amplified by a factor of $\sim$16.4 ($\sim$20.2) to $I_{\rm{S}}^{\rm{peak}}\approx 8.2 \times10^{16}$ W/cm$^2$ ($I_{\rm{S}}^{\rm{peak}}\approx 10.1\times10^{16}$ W/cm$^2$) with respect to its initial value for the Gaussian (LG) pump case.
The seed pulse significantly depletes the pump pulse, leading to a a very high pump-to-seed energy transfer efficiency, up to $\sim$56$\%$ for the Gaussian pump case and $\sim$65$\%$ for the LG pump case, respectively.

Figure \ref{fig3}(b) shows 
the spectrum of the amplified seed pulse once it has left the plasma amplifier in the LG pump case.
The final seed spectrum shows a downshifted peak at $\sim0.99\omega_0$.
This frequency shift of $\sim0.01\omega_0$ agrees well with the strong coupling theory.
The above spectrum feature confirms that the underlying mechanism is indeed SC-SBS for the whole amplification process.

We assess the robustness of the SC-SBS amplification scheme by varying several parameters. Taking the LG pump case as an example, if the initial peak seed intensity is reduced from $5\times10^{15}$ W/cm$^2$ to $1\times10^{15}$ W/cm$^2$ while the other parameters are kept unchanged, the final peak seed intensity after amplification is just reduced by $\sim$$15\%$.
Increasing both electron and ion temperatures by a factor of four will only result in a few percent difference in the amplification factor.
If the pump and seed are not perfectly head-on colliding but have a very small angle between them, its impact on our conclusions is also relatively minor.
In addition, we find that ion kinetic effects only become important after most of the seed pulse has exited the plasma. 
Therefore these kinetic effects have little influence on the SC-SBS process (see Supplementary Note 3 for details).

\subsubsection{}
\textbf{Case 2: Generation and amplification of vortex beams with new OAM modes and controllable polarization.}
In addition to amplifying vortex beams with existing OAM modes, SC-SBS also provides a mechanism for generating and amplifying vortex beams with new OAM modes and controllable polarization.
Figure \ref{fig1}(b) illustrates this process schematically.
The LG seed pulse is linearly polarized in the $x$ direction, having electric field $\vec{E_{\rm{S}}}=E_{{\rm{S}},x}\vec{e_x}\propto a_{{\rm{S}},x}\exp(il_{{\rm{S}},x}\phi )\vec{e_x}$.
The pump electric field is $\vec{E_{\rm{P}}}=E_{{\rm{P}},x}\vec{e_x}+\exp(i\psi_{\rm{P}}^{x,y})E_{{\rm{P}},y}\vec{e_y}$, which has components $E_{{\rm{P}},x}\propto a_{{\rm{P}},x}\exp(il_{{\rm{P}},x}\phi)$ and $E_{{\rm{P}},y}\propto a_{{\rm{P}},y}\exp(il_{{\rm{P}},y}\phi)$ in both $x$ and $y$ directions, respectively, with $\psi_{{\rm{P}}}^{x,y}$ the relative phase shift between $E_{{\rm{P}},x}$ and $E_{{\rm{P}},y}$. When the pump overlaps with the seed in the plasma,
an electron density perturbation as well as an IAW is initially excited due to the beating of the pump and seed electric field components in the $x$ direction.
Since $E_{{\rm{P}},x}$ carries OAM mode of $l_{{\rm{P}},x}$ and $E_{{\rm{S}},x}$ carries OAM mode of $l_{{\rm{S}},x}$, 
the electron density perturbation as well as the IAW is $\delta N \sim \exp (il_{\delta N}\phi)$ with $l_{\delta N}=l_{{\rm{P}},x}-l_{{\rm{S}},x}$ based on the $l$-number matching condition.
The same IAW also couples the pump and seed electric field components in the $y$ direction.
Therefore $l_{\delta N}=l_{{\rm{P}},y}-l_{{\rm{S}},y}$ must be satisfied in order to ensure the OAM conservation in the $y$ direction.
This implies the generation of a new seed component $E_{{\rm{S}},y}$
with OAM mode of $l_{{\rm{S}},y}=l_{{\rm{P}},y}-l_{\delta N}=l_{{\rm{P}},y}-l_{{\rm{P}},x}+l_{{\rm{S}},x}$.
Physically, it is due to the scattering of the pump component in the $y$ direction by the density perturbation [see Eq. \eqref{eq2} or Eq. \eqref{eq5}].

\begin{figure}[t]
\centering\includegraphics[height=0.65\textwidth]{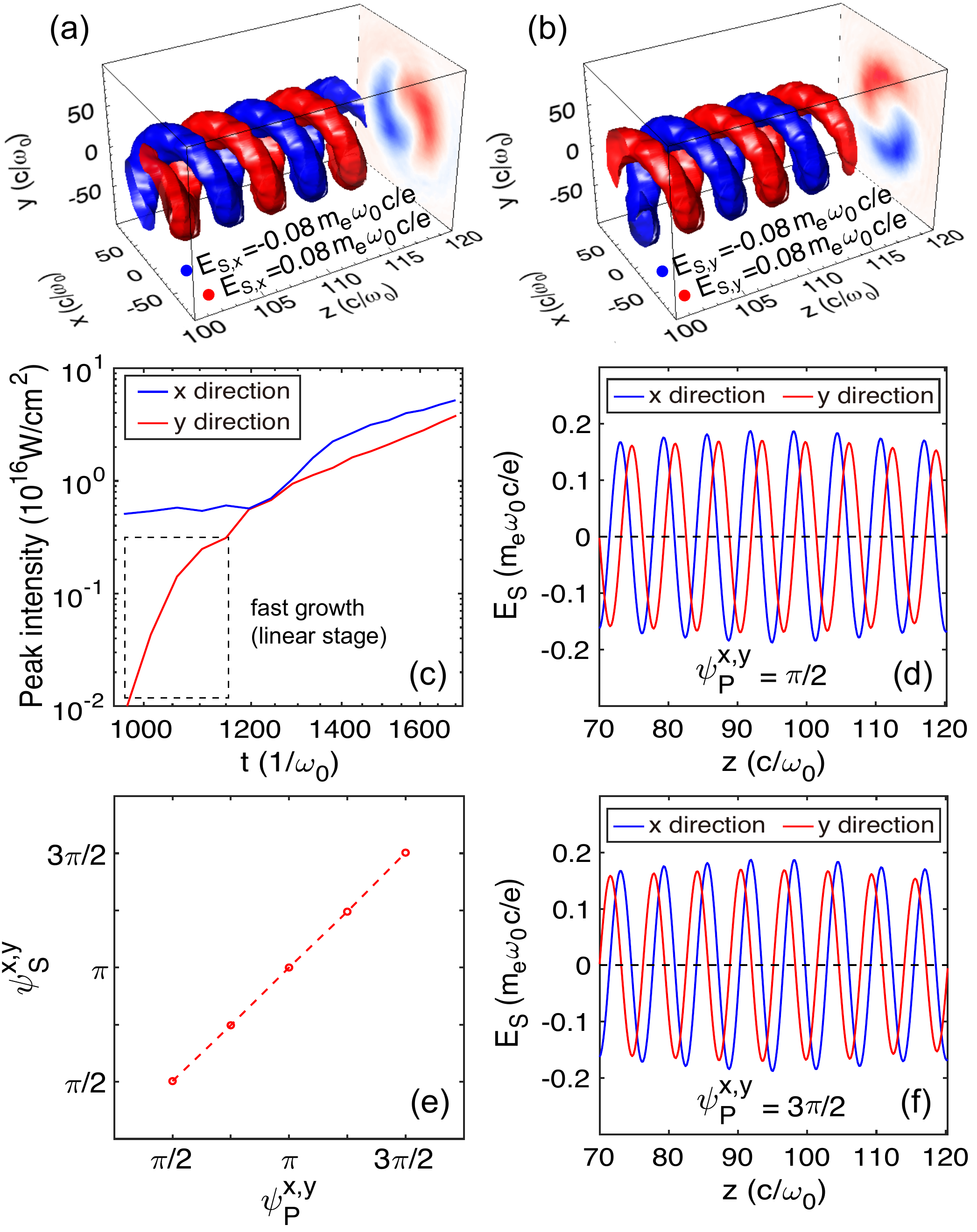}
\caption{\label{fig4} 
The 3D isosurfaces of the amplified existing component $E_{{\rm{S}},x}$ and the new component $E_{{\rm{S}},y}$ are shown in (a) and (b), respectively. Projections in the $x$-$y$ plane (blue–white–red) show the corresponding electric field slice at $z = 100c/\omega_0$.
(c) Temporal evolution of the amplified peak intensity for the two seed components. Note that both axes are presented in the logarithmic scale.
(e) The seed phase shift $\psi_{\rm{S}}^{x,y}$ vs the pump phase shift $\psi_{\rm{P}}^{x,y}$.
(d) and (f) show the lineouts of the amplified seed electric field at $t=1700 \omega_0^{-1}$, $x=55c/\omega_0$ and $y=0c/\omega_0$ in the case of $\psi_{\rm{P}}^{x,y}=\pi/2$ (d) and $\psi_{\rm{P}}^{x,y}=3\pi/2$ (f), respectively.
}
\end{figure}

By substituting the pump and seed components in both transverse directions into Eq. \eqref{eq2} or Eq. \eqref{eq5},
we can show that the relative phase shift between $E_{{\rm{S}},y}$ and $E_{{\rm{S}},x}$ is the same as that between $E_{{\rm{P}},y}$ and $E_{{\rm{P}},x}$, i.e., $\psi_{\rm{S}}^{x,y}=\psi_{\rm{P}}^{x,y}$.
We note that the relative phase shift $\psi_{\rm{P}}^{x,y}$ or $\psi_{\rm{S}}^{x,y}$ represents the polarization state (e.g., linear, circular or elliptical polarizations) of the pump or seed pulse.
Therefore, the polarization state of the amplified seed pulse ($\psi_{\rm{S}}^{x,y}$) can be easily manipulated by controlling the initial pump polarization ($\psi_{\rm{P}}^{x,y}$).

Figure \ref{fig4} shows simulation results illustrating the above mechanism.
In these simulations, the $x-$linearly polarized LG seed pulse has exactly the same parameters as those in Fig. \ref{fig2}, carrying OAM mode of $l_{{\rm{S}},x}=1$.
The pump pulse is a left-handed circularly-polarized (i.e., $a_{{\rm{P}},x}$=$a_{{\rm{P}},y}$ and $\psi_{\rm{P}}^{x,y}=\pi/2$) Gaussian beam with $l_{{\rm{P}},x}=l_{{\rm{P}},y}=0$, $W_{\rm{P}}=15\ \mu$m and $I_{{\rm{P}}0}^{\rm{peak}}=1\times10^{16}$ W/cm$^2$.
After overlapping in the plasma, the pump becomes depleted. 
For the seed, both the existing component $E_{{\rm{S}},x}$ and the generated new component $E_{{\rm{S}},y}$ are amplified.
Figures \ref{fig4}(a) and \ref{fig4}(b) show the 3D iso-surfaces of the final amplified $E_{{\rm{S}},x}$ and $E_{{\rm{S}},y}$, respectively. 
As clearly shown, the existing component $E_{{\rm{S}},x}$ featuring a single-twisted helical structure remains its initial OAM mode of $l_{{\rm{S}},x}=1$ while the new component $E_{{\rm{S}},y}$ also shows a single-twisted helical structure, which is a signature for $l_{{\rm{S}},y}=1$, in good agreement with the theoretical analysis $l_{{\rm{S}},y}=l_{{\rm{P}},y}-l_{{\rm{P}},x}+l_{{\rm{S}},x}=1$.
Field projections in the $x$-$y$ plane further confirm this conclusion.

Figure \ref{fig4}(c) illustrates the temporal evolution of the peak seed intensity in both $x$ and $y$ directions.
The amplification process of the new seed component ($l_{{\rm{S}},y}=1$ mode) first operates in the ``linear" stage ($t\lesssim1200\omega_0^{-1}$) where
the seed grows exponentially. 
For $t \gtrsim 1200\omega_0^{-1}$ the intensity of the new seed component becomes comparable to the pump intensity. 
As a result, the amplification enters the ``self-similar" stage where the pump depletion is not negligible and the seed growth slows down.
During the ``self-similar" stage, both the existing and the new seed components amplify 
at roughly the same growth rates,
which is in agreement with theory. 
Since it grows from initially higher peak intensity, the existing $l_{{\rm{S}},x}=1$ mode reaches a slightly higher final intensity ($I_{{\rm{S}},x}^{\rm{peak}}=5\times10^{16}$ W/cm$^2$) than the new $l_{{\rm{S}},y}=1$ mode ($I_{{\rm{S}},y}^{\rm{peak}}=4\times10^{16}$ W/cm$^2$). 
Thus the total peak intensity of the seed is $9\times10^{16}$ W/cm$^2$, corresponding to an amplification factor of $\sim$18 with respect to its initial value of $I_{{\rm{S}}0}^{\rm{peak}}=5\times10^{15}$ W/cm$^2$.

Figure \ref{fig4}(d) shows the electric field waveforms of the amplified $E_{{\rm{S}},x}$ and $E_{{\rm{S}},y}$.
It is apparent that the simulated phase shift $\psi_{\rm{S}}^{x,y}$ between the final $E_{{\rm{S}},x}$ and $E_{{\rm{S}},y}$ is approximately $\pi/2$, 
which agrees well with the theoretical result $\psi_{\rm{S}}^{x,y}=\psi_{\rm{P}}^{x,y}$. 
The amplitude ratio $\epsilon_{\rm{S}}$ between the two electric components is $\epsilon_{\rm{S}} = \min \left(a_{{\rm{S}},x}^{\rm{peak}}, a_{{\rm{S}},y}^{\rm{peak}}\right)/\max \left(a_{{\rm{S}},x}^{\rm{peak}}, a_{{\rm{S}},y}^{\rm{peak}}\right)=\min \left(\sqrt{I_{{\rm{S}},x}^{\rm{peak}}}, \sqrt{I_{{\rm{S}},y}^{\rm{peak}}}\right)/\max \left(\sqrt{I_{{\rm{S}},x}^{\rm{peak}}}, \sqrt{I_{{\rm{S}},y}^{\rm{peak}}}\right)\approx 0.9$.
Considering the amplitude ratio close to unity, together with $\psi_{\rm{S}}^{x,y}=\pi/2$,
the final seed is left-handed quasi-circularly polarized. 
When the polarization state of the pump beam changes, the output seed polarization will also change accordingly.
For example, in the simulations,
by continuously changing $\psi_{\rm{P}}^{x,y}$ from $\pi/2$ to $3\pi/2$, 
$\psi_{\rm{S}}^{x,y}$ also changes from $\pi/2$ to $3\pi/2$ [Fig. \ref{fig4}(e)] while the amplitudes of the two electric field components remain constant.
Therefore, the polarization state of the seed can be tuned from left-handed quasi-circular through elliptical and linear to an elliptical polarization of opposite helicity.
Especially, when $\psi_{\rm{P}}^{x,y}=3\pi/2$, the output seed is quasi-circularly polarized but with opposite handedness (right-handed) [Fig. \ref{fig4}(f)].
This shows the feasibility of reversing the rotation direction of the amplified seed pulse simply by switching the handedness of the pump laser.

\subsubsection{}
\textbf{Case 3: Generation and amplification of vector beams from vortex seed beams.}
Apart from the LG vortex beam, SC-SBS process can also be used to generate and amplify vector laser beams with space-variant polarization states.
The simplest example of a vector beam is a radially or azimuthally polarized cylindrical vector beam, 
which can be viewed as a linear combination of LG vortex beams with orthogonal circular polarizations and different OAM modes of 1 and -1. 
By considering the superposition of even higher-order OAM modes, 
even more exotic vector beams (e.g., having spider- and web-like polarizations) \cite{Light_Optical_Vorticies, Khajavi_2016} can be generated. 

\begin{figure}[h]
\centering\includegraphics[height=0.64\textwidth]{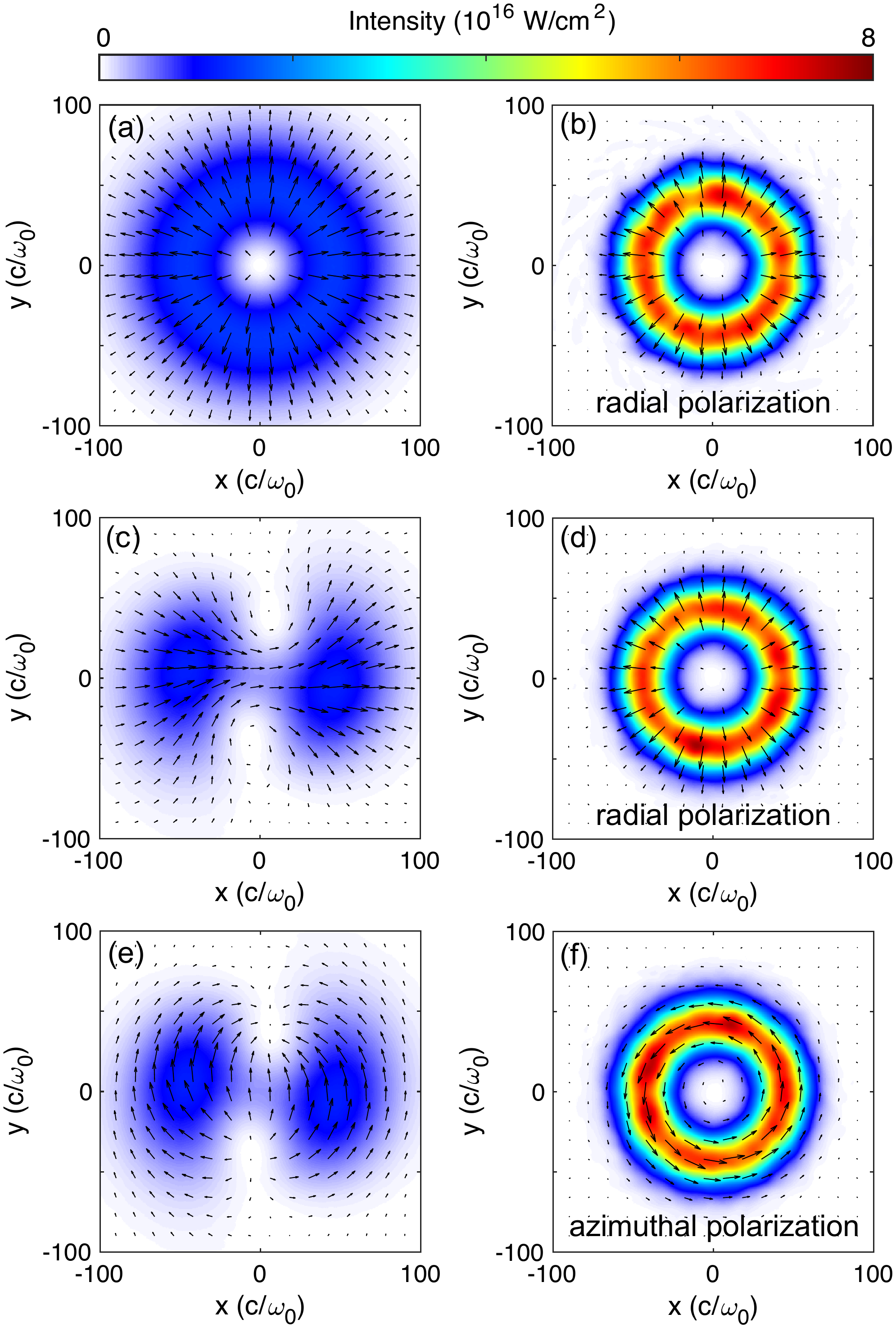}
\caption{\label{fig5} 
Spatial distributions of intensities and polarization vectors for the incident pump (left column) and final amplified seed (right column) at a given $z$.
The top row shows that the interaction between a radially-polarized vector pump ($\psi_{\rm{P}}^{\rm{R,L}}=0$, $l_{\rm{P,R}}=1$, $l_{\rm{P,L}}=-1$) and a vortex seed with 
$l_{\rm{S,L}}=-1$
leads to the generation and amplification of a quasi-radially polarized intense pulse.
The middle row shows that by changing the above pump to a pulse ($\psi_{\rm{P}}^{\rm{R,L}}=0$, $l_{\rm{P,R}}=2$, $l_{\rm{P,L}}=0$, $W_{0,\rm{P,R}}=8\ \mu$m and $W_{0,\rm{P,L}}=15\ \mu$m) without radial polarization while keeping the seed unchanged, we can still obtain a quasi-radially polarized intense pulse.
The bottom row shows that compared with the middle row, if we only change the phase shift $\psi_{\rm{P}}^{\rm{R,L}}$ from 0 to $\pi$, the final amplified seed will be a quasi-azimuthally polarized vector beam.
}
\end{figure}

The process is shown schematically in Fig. \ref{fig1}(c).
The seed pulse is a circularly polarized (assuming left-handed) LG vortex beam, having electric field $\vec{E_{\rm{S}}}=E_{\rm{S,L}}\vec{e_{\rm{L}}}\propto a_{\rm{S,L}}\exp(il_{\rm{S,L}}\phi )\vec{e_{\rm{L}}}$.
The pump electric field is $\vec{E_{\rm{P}}}=E_{\rm{P,R}}\vec{e_{\rm{R}}}+\exp(i\psi_{\rm{P}}^{\rm{R,L}})E_{\rm{P,L}}\vec{e_{\rm{L}}}$, which has components $E_{\rm{P,R}}\propto a_{\rm{P,R}}\exp(il_{\rm{P,R}}\phi)$ and $E_{\rm{P,L}}\propto a_{\rm{P,L}}\exp(il_{\rm{P,L}}\phi)$ in both right-handed ($\vec{e_{\rm{R}}}$) and left-handed ($\vec{e_{\rm{L}}}$) directions, respectively, with $\psi_{\rm{P}}^{\rm{R,L}}$ the relative phase shift between $E_{\rm{P,R}}$ and $E_{\rm{P,L}}$.
Similar to the linear polarization cases, the OAM selection rules also hold for orthogonal circular polarization cases.
Therefore, when the seed carrying OAM mode of 
$l_{\rm{S,L}}$
interacts with the pump carrying OAM modes of $l_{\rm{P,R}}$ and $l_{\rm{P,L}}$ in the plasma,
a new right-handed circularly polarized seed component $E_{\rm{S,R}}$ 
will be generated so that OAM is conserved in both left-handed and right-handed directions. 
The OAM of this new seed component is 
$l_{\rm{S,R}}=l_{\rm{P,R}}-l_{\rm{P,L}}+l_{\rm{S,L}}$.
In addition, the phase shift between the two seed components is identical to that of the pump pulse, i.e., $\psi_{\rm{S}}^{\rm{R,L}}=\psi_{\rm{P}}^{\rm{R,L}}$.
When $l_{\rm{P,L}}\neq l_{\rm{P,R}}$, we can obtain $l_{\rm{S,L}}\neq l_{\rm{S,R}}$, that is, the amplified seed pulse has two different OAM modes in two orthogonal circular directions.
Superposition of these two modes then leads to the generation of a vector beam.

To examine the above predictions, we present simulation results in Figs. \ref{fig5}(a) and \ref{fig5}(b), 
where a left-handed circularly polarized LG seed pulse with $l_{\rm{S,L}}=-1$,
$W_{0,\rm{S}}=10\ \mu$m and $I_{\rm{S}0}^{\rm{peak}}=1\times10^{16}$ W/cm$^2$ is Brillouin amplified by a 
pump beam with $a_{\rm{P}0,\rm{R}}$=$a_{\rm{P}0,\rm{L}}$, $\psi_{\rm{P}}^{\rm{R,L}}=0$, $l_{\rm{P,R}}=1$, $L_{\rm{P,L}}=-1$, $W_{0,\rm{P}}=10\ \mu$m and $I_{\rm{P}0}^{\rm{peak}}=1\times10^{16}$ W/cm$^2$.
Indeed, such a pump beam composed of opposite circularly polarized LG modes with $l_{\rm{P,R}}=1$ and $l_{\rm{P,L}}=-1$ is exactly a radially polarized vector beam [Fig. \ref{fig5}(a)]\cite{Light_Optical_Vorticies}.
After interaction in the plasma, 
a new seed component $E_{\rm{S,R}}$ with right-handed circular polarization and OAM mode of $l_{\rm{S,R}}=l_{\rm{P,R}}-l_{\rm{P,L}}+l_{\rm{S,L}}=1$ is generated. 
Finally, both the existing component $E_{\rm{S,L}}$ and the generated new component $E_{\rm{S,R}}$ 
are amplified to roughly the same intensity,
resulting in a total peak seed intensity of $\sim$$8\times10^{16}$ W/cm$^2$ and an amplification factor of $\sim$16.
By superposing these two seed components together, the total amplified pulse will also be a quasi-radially polarized vector beam, as shown in Fig. \ref{fig5}(b).
This also satisfies the polarization matching rule, where the output seed polarization is determined by the input pump polarization.

In addition to directly using a radially polarized pump,
it is also possible to generate and amplify a radially polarized seed via SC-SBS from a pump without radial polarization.
An example is shown in Figs. \ref{fig5}(c) and \ref{fig5}(d), where simulations demonstrate that
if we change $l_{\rm{P,R}}$ from 1 to 2 and $l_{\rm{P,L}}$ from -1 to 0 (Gaussian mode),
we can still obtain an amplified quasi-radially polarized seed pulse [Fig. \ref{fig5}(d)].
The reason is that, although $l_{\rm{P,R}}$ and $l_{\rm{P,L}}$ are changed, the difference value between $l_{\rm{P,R}}$ and $l_{\rm{P,L}}$ remains unchanged.
Thus the generated new right-handed circularly polarized seed component still has OAM mode of $l_{\rm{S,R}}=l_{\rm{P,R}}-l_{\rm{P,L}}+l_{\rm{S,L}}=1$.
Furthermore, 
if we change the phase shift $\psi_{\rm{P}}^{\rm{R,L}}$ between the two pump components from $\psi_{\rm{P}}^{\rm{R,L}}=0$ to $\psi_{\rm{P}}^{\rm{R,L}}=\pi$, then the phase shift $\psi_{\rm{S}}^{\rm{R,L}}$ between the two seed components will also change accordingly to $\psi_{\rm{S}}^{\rm{R,L}}=\pi$ and the final amplified seed will be a quasi-azimuthally polarized vector beam, as shown in Figs. \ref{fig5}(e) and \ref{fig5}(f).
Apart from radial and azimuthal polarizations, SC-SBS can create even more exotic vectorial polarizations and amplify them to very high intensities (see Supplementary Note 4 for examples of spider- and web-like polarizations).

\subsubsection{}
\textbf{Case 4: Generation and amplification of vortex beams from vector seed beams.}
The SC-SBS can not only couple SAM and OAM to generate vector beams from vortex seeds, but the converse is also true. It can decouple SAM and OAM to generate vortex beams from vector seeds. Fig. \ref{fig1}(d) shows this process. 
The seed pulse is a vector beam composed of two opposite circularly polarized LG modes with different OAM modes, having electric field $\vec{E_{\rm{S}}}=E_{\rm{S,R}}\vec{e_{\rm{R}}}+\exp(i\psi_{\rm{S}}^{\rm{R,L}})E_{\rm{S,L}}\vec{e_{\rm{L}}}$ with $E_{\rm{S,R}}\propto a_{\rm{S,R}}\exp(il_{\rm{S,R}}\phi)$ and $E_{\rm{S,L}}\propto a_{\rm{S,L}}\exp(il_{\rm{S,L}}\phi)$.
The pump pulse is a circularly polarized (assuming left-handed) Gaussian ($l_{\rm{P,L}}=0$) or LG vortex ($l_{\rm{P,L}}\neq0$) beam with $\vec{E_{\rm{P}}}=E_{\rm{P,L}}\vec{e_{\rm{L}}}\propto a_{\rm{P,L}}\exp(il_{\rm{P,L}}\phi )\vec{e_{\rm{L}}}$.
During the interaction in the plasma, 
the scattering of the pump by the density perturbation leads to the energy delivery from the pump to the seed,
thus only the seed component which has the same circular polarization handedness as the pump, i.e., 
$E_{\rm{S,L}}$, can be ``selected" and amplified.
The other seed component with opposite polarization handedness, i.e., $E_{\rm{S,R}}$, will not be amplified. 
Finally, the amplified seed pulse becomes a quasi-vortex beam carrying OAM mode of 
$l_{\rm{S,L}}$
when the amplification factor of $E_{\rm{S,L}}$ is high.

\begin{figure}[t]
\centering\includegraphics[height=0.44\textwidth]{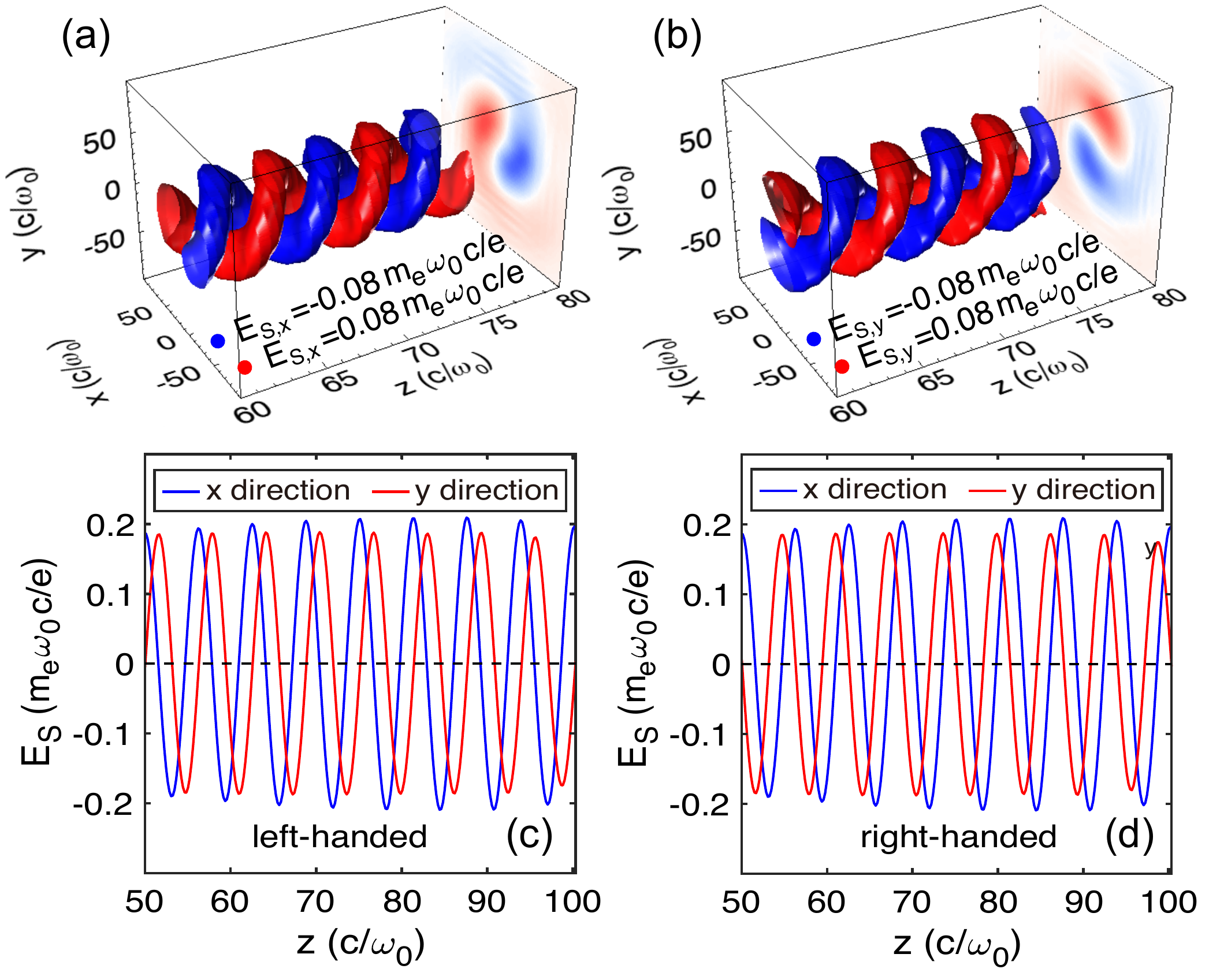}
\caption{\label{fig7} 
Simulations results showing the generation and amplification of vortex beams from azimuthally polarized vector seeds by left-handed (a-c) or right-handed (d) circularly polarized Gaussian pumps.
(a) and (b) show the 3D isosurfaces of the amplified seed electric fields in the $x$ and $y$ directions, respectively. Projections in the $x$-$y$ plane (blue–white–red) show the corresponding electric field slice at $z = 60c/\omega_0$.
(c) and (d) show the lineouts of the amplified seed electric field at $t=1700 \omega_0^{-1}$, $x=66c/\omega_0$ and $y=0c/\omega_0$ in the left-handed (c) and right-handed (d) circular pump polarization cases, respectively.
}
\end{figure}

Figure \ref{fig7} shows simulation results confirming the above prediction.
In the simulation, the pump is a left-handed circularly polarized Gaussian beam ($l_{\rm{P,L}}=0$) 
with $W_{0,\rm{P}}=14\ \mu$m and $I_{\rm{P}0}^{\rm{peak}}=1\times10^{16}$ W/cm$^2$.
The seed pulse is an azimuthally polarized vector beam with $W_{0,\rm{S}}=12\ \mu$m and $I_{\rm{S}0}^{\rm{peak}}=5\times10^{15}$ W/cm$^2$, which is a superposition of right-handed circular polarized LG mode with $l_{\rm{S,R}}=1$ and a left-handed circular polarized LG mode with $l_{\rm{S,L}}=-1$. 
The relative phase shift between these two seed modes is $\psi_{\rm{P}}^{\rm{R,L}}=\pi$.
Note that there is no initial OAM in this configuration, since both the pump ($l_{\rm{P,L}}=0$) and seed ($l_{\rm{S,R}}+l_{\rm{S,L}}=0$) have no net OAM.
Simulation results show that the plasma can serve as an OAM mode selector and amplifier during the SC-SBS process.
It results in the generation and amplification of a left-handed quasi-circularly-polarized OAM mode with $l_{\rm{S,L}}=-1$, as shown in Figs. \ref{fig7}(a)-\ref{fig7}(c).
The final peak intensity reaches $\sim$$8.6\times10^{16}$ W/cm$^2$, corresponding to an amplification factor of $\sim$17.2 with respect to its initial value.
In addition, if we change the circular polarization handedness of the pump beam from left-handed to right-handed while keep the other parameters
constant,
the polarization state of the output seed beam will also undergo a similar change.
Finally, we can obtain a right-handed quasi-circularly polarized amplified pulse with $l_{\rm{S,R}}=1$, as shown in Fig. \ref{fig7}(d).
These results not only confirm the polarization matching rule between the pump and the seed, but also show that SC-SBS can provide a mechanism for the production and amplification of vortex beams carrying OAM from a configuration with no net OAM.

\section*{Discussion}
In most of our simulation cases, 
the pulse duration (FWHM) of the amplified pulse is compressed from initial 100 fs to final $\sim$60 fs.
The amount of energy transferred from the pump to the seed reaches as high as $\sim 65\%$ of the total pump energy.
This efficiency is much higher than previous Raman amplification schemes \cite{NC_Vieira}.

Due to limited computation resources, we use short plasma lengths (80 $\mu$m), high plasma densities (0.3$n_{\rm{c}}$) and moderate pump/seed spot sizes ($\sim$10 $\mu$m scale) in the simulations to reduce the simulation box size and simulation time.
In this case, the simulated seed intensities after amplification can reach as high as $\sim10^{17}$ W/cm$^2$, leading to final peak powers of $\sim$TW level.
However, this power is not the upper limit.
Recently, both theoretical studies via 2D PIC simulations \cite{PRL_Weber, Alves_2021} and experimental studies \cite{PRX_experiment} have shown that SC-SBS can support much larger spot sizes ($\sim$100 $\mu$m) for Gaussian beams and amplify them to even hundreds of TW level by controlling the competing filamentation or SRS (occurs when $n_{\rm{e0}}<0.25n_{\rm{c}}$) instabilities through optimization strategies (e.g, using lower plasma density and shaping the plasma density profile \cite{PRL_Weber}).
Similarly, 
provided that the pump and seed spot sizes can be increased to $\sim$100 $\mu$m scale without affecting the final intensity, our simulation results suggest that SC-SBS can also be used to generate and amplify vortex and vector laser beams to $\sim$100 TW level.
In addition, 
if some advanced pump focusing schemes such as the ``flying focus" technique \cite{NP_Flying_focus, PRL_FF_Raman} are utilized 
to further mitigate spontaneous pump instabilities before meeting the seed, this would help further push amplification and also preserve the quality (high contrast) of these exotic pulses at high intensities.

In summary, we have proposed an efficient scheme that can generate and amplify vortex and vector beams to high intensities/powers by employing plasma-based SC-SBS.
We note that although Vieira et al. have recently shown the matching of OAM and generation/amplification of vortex beams in plasma-based SRS \cite{NC_Vieira}, they have not considered manipulation of polarization states and generation/amplification of vector beams. 
In contrast, 
we have exploited in detail both OAM selection rules and polarization matching properties of SC-SBS in plasmas.
Our results show that it is possible to simultaneously manipulate the OAM and polarization states of vortex pulses while amplifying them during the SC-SBS process.
We have also investigated various couplings between the pump and the seed, 
demonstrating how OAM and SAM of 
them can be manipulated to create exotic vector pulses.
This scheme paves the way for novel optical devices such as plasma-based amplifiers and manipulators for structured light pulses.



\section*{Methods}
\textbf{Numerical Simulation Setup.}
We have performed 3D PIC simulations using the code OSIRIS \cite{Fonseca_2002}. Osiris is a fully parallel, fully relativistic PIC code that solves the full set of Maxwell’s equations on a grid using currents and charge densities calculated by weighting discrete particles onto the grid. Each particle is then pushed to a new position and momentum via self-consistently calculated fields.
Thus, OSIRIS makes no physical approximations to the extent where quantum mechanical effects can be neglected.

The dimensions of the 3D simulation box are 660$\times$350$\times$350$(c/\omega_0)^3$ (105$\times$55.7$\times$55.7$\mu$m$^3$) for all the simulations presented in the paper, divided by 4000$\times$800$\times$800 cells along the $z$, $x$ and $y$ directions, respectively. 
This corresponds to 38 grid cells per pump/seed wavelength in the $z$ direction, sufficient to resolve the SC-SBS physical process.
8 macroparticles per cell ($N_{\rm{ppc}}=8$) are used for both the plasma electrons and ions ($\sim$4$\times 10^{10}$ macroparticles in total).
Note that the number of macroparticles per cell $N_{\rm{ppc}}$ is not large enough due to limited computation resource, 
therefore the numerical thermal noise level (depending on $1/N_{\rm{ppc}}$) in the simulation is larger than the real thermal noise in a plasma.
This will not influence the SC-SBS process since the energy transfer is controlled by the seed pulse and not by the noise.
However, the competing thermal filamentation instability of the pump is likely to be overestimated. 
In this sense the PIC approach here represents a worst-case scenario. 

\section* {Acknowledgments}
This work was supported by the U.S. Department of Energy Grant No. DE-SC0010064 and NSF Grant No. 2003354. This research used computing resources of the National Energy Research Scientific Computing Center (NERSC), a U.S. Department of Energy Office of Science User Facility located at Lawrence Berkeley National Laboratory, operated under Contract No. DE-AC02-05CH11231.

\section* {Data availability}
All relevant data are available from the corresponding authors upon reasonable request.

\section* {Code availability}
The codes that support the plots within this paper and other findings of this study are available from the corresponding authors upon reasonable request.

\section* {Author contributions}
Y. W. conceived and developed the main idea. Y. W. conducted theoretical calculations and simulations. Y. W. and C. J. wrote the manuscript. All authors commented on the manuscript.

\section* {Competing interests}
The authors declare no competing interests.

\section*{References}
\bibliography{refs_SBS}   

\end{document}